\documentclass[aoas,MSNbibl,nameyear,dvips]{arximspdf}
\usepackage{graphicx}
\doi{10.1214/13-AOAS649} 
\volume{7}
\issue{3}
\pubyear{2013}
\firstpage{1540}
\lastpage{1561}

\makeatletter

\newcommand{\eqref}[1]{(\ref{#1})}
\newcommand{\bW}{\mathbf{W}}
\newcommand{\bX}{\mathbf{X}}
\newcommand{\bu}{\mathbf{u}}
\newcommand{\bv}{\mathbf{v}}
\newcommand{\bK}{\mathbf{K}}
\newcommand{\cH}{\mathcal{H}}
\newcommand{\cW}{\mathcal{W}}
\newcommand{\cU}{\mathcal{U}}
\newcommand{\cV}{\mathcal{V}}
\newcommand{\cY}{\mathcal{Y}}
\newcommand{\cZ}{\mathcal{Z}}
\newcommand{\bbR}{\mathbb{R}}
\newcommand{\bOmega}{\bolds{\Omega}}
\newcommand{\diag}{\operatorname{diag}}

\newproclaim{definition}{Definition}
\makeatother

\begin{document}
\begin{frontmatter}

\title{Robust regularized singular value decomposition with application
to mortality data}
\runtitle{Robust regularized singular value decomposition}

\begin{aug}
\author[a]{\fnms{Lingsong}~\snm{Zhang}\corref{}\ead[label=e1]{lingsong@purdue.edu}},
\author[b]{\fnms{Haipeng} \snm{Shen}\ead[label=e2]{haipeng@email.unc.edu}\thanksref{t2}}
\and
\author[c]{\fnms{Jianhua Z.} \snm{Huang}\ead[label=e3]{jianhua@stat.tamu.edu}\thanksref{t3}}
\thankstext{t2}{Supported in part by NIH/NIDA (1 RC1 DA029425-01) and
NSF (CMMI-0800575, DMS-11-06912).}
\thankstext{t3}{Supported in part by NCI (CA57030), NSF (DMS-09-07170,
DMS-10-07618, DMS-12-08952) and Award No. KUS-C1-016-04,
made by King Abdullah University of Science and Technology (KAUST).}
\affiliation{Purdue University, University of North Carolina and\break Texas
A\&M University}
\address[a]{L. Zhang\\
Department of Statistics\\
Purdue University\\
150 N. University St.\\
West Lafayette, Indiana 47906\\
USA\\
\printead{e1}}
\address[b]{H. Shen\\
Department of Statistics and Operations Research\\
University of North Carolina\\
Chapel Hill, North Carolina 27599\\
USA\\
\printead{e2}}
\address[c]{J. Z. Huang\\
Department of Statistics \\
Texas A\&M University\\
3143 TAMU\\
College Station, Texas 77843-3143 \\
USA\\
\printead{e3}}
\runauthor{L. Zhang, H. Shen and J. Z. Huang}
\end{aug}

\received{\smonth{10} \syear{2012}}
\revised{\smonth{2} \syear{2013}}

%
\begin{abstract}
We develop a robust regularized singular value decomposition (\mbox{RobRSVD})
method for analyzing two-way functional data. The research is motivated
by the application of modeling human mortality as a smooth two-way
function of age group and year. The RobRSVD is formulated as a
penalized loss minimization problem where a robust loss function is
used to measure the reconstruction error of a low-rank matrix
approximation of the data, and an appropriately defined two-way
roughness penalty function is used to ensure smoothness along each of
the two functional domains. By viewing the minimization problem as two
conditional regularized robust regressions, we develop a fast iterative
reweighted least squares algorithm to implement the method. Our
implementation naturally incorporates missing values. Furthermore, our
formulation allows rigorous derivation of leave-one-row/column-out
cross-validation and generalized cross-validation criteria, which
enable computationally efficient data-driven penalty parameter
selection. The advantages of the new robust method over nonrobust ones
are shown via extensive simulation studies and the mortality rate application.
\end{abstract}

%

\end{frontmatter}

\section{Introduction}

This paper develops a \textit{robust} regularized singular value
decomposition (SVD) method for \textit{two-way functional} data.
One-way functional data analysis (FDA) focuses on a population of
curves or functions and has gained much attention in the last
decade or so, as well documented in \citeauthor{ramssilv02}
(\citeyear{ramssilv02,ramssilv05})
and \citet{Ferraty2006NonparametricFDA}. Different from one-way
functional data, two-way functional data are functions in two ways:
both index domains $I$ and $J$ of the data matrix
$\mathbf{X}= (x_{i,j} )_{i\in I, j\in J}$ are structured with
notions of smoothness, that is, both rows and columns of the data matrix
can be viewed as discretizations of some underlying smooth functions
[\citet{huang2009analysis}]. For example, in our motivating Spanish
mortality application (Section~\ref{spaindataanalysis}), the data
matrix records mortality rates for different age groups between
ages 0 and 110 (columns) in Spain from year 1908 to 2007 (rows).
It is reasonable to
consider the mortality rate as a smooth function of both age and time
period. Similar two-way functional structure also exists in many other
applications. For example, the network traffic pattern
in \citet{zhang2007singular} is a smooth function of
time-of-the-day and calendar date; the call center customer patience
in \citet{huang2009analysis} is a smooth function of customer waiting
time and time-of-the-day; and the magnetoencephalography signal
in \citet{tian2011meg} is a smooth function of signal recording time
and brain spatial location.

Recently, \citet{huang2009analysis} proposed a regularized singular
value decomposition (RSVD) method for dimension reduction and feature
extraction of two-way functional data. It is based on minimization
of a regularized sum of squared reconstruction errors of a low-rank matrix
approximation. Since the squared-error loss function is used to
measure the size of reconstruction errors, the results of applying
the RSVD are sensitive to outliers. Outliers in two-way functional
data can appear in various forms, such as outlying cells, columns,
rows or blocks
(Section~\ref{SimuSec}). For example, the Spanish mortality data
contain two outlying time periods and, as we will demonstrate in
Section~\ref{spaindataanalysis}, they significantly affect the
estimation of the underlying smooth mortality trend across year
when applying the RSVD. One major contribution of the current paper is
to develop a robust regularized SVD method that can mitigate outlying
effects in two-way functional data analysis, which, to the best of our
knowledge, is the first of its kind.

To give some background on our proposed method for two-way functional
data, we first review several relevant robust functional principal component
analysis (PCA) methods that have been developed for analyzing one-way
functional data. \citet{locamarr99} proposed a robust PCA approach,
which projects the data onto a sphere or an ellipse around a robust
estimate of the center of the data, and then performs the usual PCA on the
covariance matrix of the projected data. \citet{gervini2008robust}
extended the approach of \citet{locamarr99} to functional data,
introduced the concepts of functional median and functional spherical
principal components (PC), and established the corresponding
robustness properties of the approach. \citet{hyndman2007robust} and
\citet{hyndman2009forecasting} used a projection pursuit (PP) approach
for robust functional PCA; \citet{boente2010robust} recently studied
the asymptotic robustness properties of this PP approach in terms of
influence function and breakdown point. On the other
hand, \citet{bai2008supervised} proposed a supervised SVD technique,
which can be combined with independent component analysis to improve
the robustness of analyzing functional MRI brain images.
\citet{gervini2009detecting} considered irregularly and sparsely
sampled functional data, used basis expansions to model the
functional trajectories, and modeled the functional PC scores and the
reconstruction errors using heavy-tailed distributions such as $t$ or
Cauchy to achieve robustness. All this work has focused on one-way
functional data.

We now introduce some notation to facilitate the discussion of
our proposed robust regularized SVD method for two-way functional
data. Sometimes it is reasonable to use the term---functional SVD---instead of regularized SVD to emphasize the focus on
functional data. We view the
element $x_{ij}$ of the $m\times n$ data matrix $\mathbf{X}$ as
evaluation of
an underlying \textit{smooth} function $X(\cdot,\cdot)$ on a rectangular
grid of sampling points $(y_i, z_j)$, where $y_i\ (i = 1, \ldots, m)$
are from a domain $\mathcal{Y}$ and $z_j\ (j = 1,\ldots,n)$ are from a
domain $\mathcal{Z}$. According to \citet{huang2009analysis}, the RSVD
for two-way functional data can be considered as fitting the following
smooth rank-$r$ approximation model for the two-way functional data:
%
%
\begin{equation}
\label{eqfunc-svd} X(y,z) = U_1(y) V_1(z) +
U_2 (y) V_2(z) + \cdots+ U_r(y)
V_r(z) + \varepsilon(y,z),
\end{equation}
where $U_k(y)$ and $V_k(z)$ are smooth functions on their respective domains,
and $\varepsilon(y,z)$ is a mean zero random noise.
Model \eqref{eqfunc-svd} can be thought of
as a truncated version of the singular value decomposition of bivariate
functions [\citet{gervini10}], and the orthonormal constraints
$\int V_k(z)V_l(z) \,dz=\delta_{kl}$, where $\delta$ is the
Kronecker delta, are usually imposed for identifiability.
The low-rank approximation formulation indicates that the proposed SVD
method is useful for dimensionality reduction and feature selection.
The smoothness requirement on $U_k(y)$ and $V_k(z)$ takes into account
the underlying continuity of the functional data. It is important to
note that the
SVD formulation offers a symmetric treatment of the two domains. The
existing robust functional PCA methods cannot be directly extended to
two-way functional data, because PCA treats the rows and the columns
asymmetrically. We are therefore led to the SVD which offers symmetric
treatment.

To give a simple description of our approach, we focus on extracting
the first pair of components in \eqref{eqfunc-svd}, $U_1(y)$ and
$V_1(z)$, whose discretized realizations are, respectively, denoted as
$\mathbf{u}_1\equiv(U_1(y_1), \ldots, U_1(y_m) )^T$ and $\mathbf
{v}_1\equiv(V_1(z_1), \ldots,  V_1(z_n) )^T$. Subsequent
pairs are extracted sequentially after removing the effects of the
preceding pairs. This sequential approach allows the different pairs
of components to have differing smoothness. The extracted components
should possess two desirable features---smoothness and robustness
against outliers.
We propose to solve the following problem:
%
%
\begin{equation}
\label{eqrobust} (\mathbf{u}_1, \mathbf{v}_1)\equiv
\mathop{\operatorname{argmin}}_{\mathbf{u}, \mathbf
{v}} \bigl\{\rho\bigl(\mathbf{X}-\mathbf{u}
\mathbf{v}^T\bigr)+\mathcal{P}_\lambda(\mathbf{u}, \mathbf{v})
\bigr\},
\end{equation}
where $ \mathbf{u}$ and $ \mathbf{v}$ are $m$-dimensional and $n$-dimensional
vectors, respectively, $\rho(\cdot)$ is a robust loss function,
$\mathcal{P}_{\bolds{\lambda}}(\mathbf{u}, \mathbf{v})$ is a two-way
roughness penalty to ensure smoothness for the $\mathbf{u}$ and
${\mathbf{v}}$, and $\bolds{\lambda}$ is a vector of penalty parameters.

This formulation is very general, allowing the flexibility in
the choice of the loss function and the penalty function.
Although various robust loss functions in the robust statistics
literature [\citet{huberRonchetti09}] can be used in our framework, we
focus on a typical Huber's function for its easy implementation and
fast computation. If the nonrobust squared-error loss is used,
then the penalized criterion function in \eqref{eqrobust} reduces
to the minimizing criterion for the RSVD of
\citet{huang2009analysis}. By using a robust loss function, our framework
essentially robustifies the RSVD method and, therefore, we refer to our
approach as robust regularized SVD, or \textit{RobRSVD} for short.
On the other hand, without the penalty term, the criterion in \eqref{eqrobust}
offers another way for robust SVD [\citet{amma93,liuhawk03}]; hence,
RobRSVD can also be interpreted
as smoothing of a robust SVD.

In this paper, we adopt the two-way roughness penalty function
introduced in \citet{huang2009analysis}, which has several desirable
properties for two-way regularization. Other choices of penalty
functions are possible such as the ones that shrink the functional
components to certain subspaces, for example, spaces of periodic
functions. Our framework also offers one-way robust functional
data analysis as a special case if one only imposes
roughness penalty on one of the functional domains such as
the one that corresponds to the row or the column of the data matrix.
One important feature of our method is that it works directly
with the raw observed data; there is no need to pre-smooth the
raw data, nor to obtain a robust estimate of the high-dimensional
covariance matrix, which can be computationally challenging for
one-way functional data and even more technically difficult for two-way
functional data.

We develop an
efficient iterative reweighted least squares (IRLS) algorithm
to solve the minimization problem \eqref{eqrobust}. Our algorithm
iteratively updates $\mathbf{u}$ and $\mathbf{v}$ conditioning
on the other, where each updating
step can be viewed as a (regularized) robust regression. This
view of \eqref{eqrobust} as conditional robust regressions
suggests that many robust
regression procedures can be used, such as the M-estimator
[\citet{huberRonchetti09}], the $L_1$ estimator [\citet{croufilz03}],
the least median of squares (LMS) and the least trimmed squares (LTS)
estimators [\citet{rous84}], and the IRLS estimator
[\citet{heibbeck92}]. We choose the IRLS estimator in this paper for
the following two reasons. First, it enables us to interpret the
conditional regularized robust regressions as regularized weighted
least squares. Based on this interpretation, we can rigorously derive
explicit shortcut formula for leave-one-row/column-out cross-validation
and related generalized cross-validation (GCV) scores; hence,
data-driven selection of the penalty parameters can be carried out very
efficiently. Note that the selection of the penalty parameters for the
row and column is naturally decoupled due to the conditional regression
perspective. Second, the IRLS estimator is used due to its fast
computation and comparable performance when compared against several
other robust regression procedures, as shown by 
\citet{shen2007robust}. The alternating estimation procedure also
suggests a natural way to incorporate missing values.

The remainder of the paper is organized as follows.
Section~\ref{RobFSVDmethod} describes
technical details of the RobRSVD method, including formulation,
the IRLS algorithm, penalty parameter selection, treatment of missing
values and interpolation of results in function space.
Results of simulation studies are presented in Section~\ref{SimuSec} to
compare the performance of RobRSVD with standard SVD and the
regularized SVD (RSVD) of \citet{huang2009analysis}. Section~\ref{spaindataanalysis} analyzes the motivating Spanish mortality
application and demonstrates the practical advantages of \mbox{RobRSVD} over
the other two methods.

\section{The methodology} \label{RobFSVDmethod}

We describe the RobRSVD method in this section.
Section~\ref{secformulation} gives
its formulation, Section~\ref{IRLSSec} derives the IRLS algorithm, and
Sections~\ref{GCVSec}--\ref{secRKHS} discuss several
implementation details. 

\subsection{Formulation}\label{secformulation}
It is well known that the SVD can be viewed as finding a sequence of
rank-one matrix approximations of a data matrix [\citet{gabrzami79}].
We adapt this
idea to define the RobRSVD as a method for obtaining a sequence of
\textit{robust regularized} rank-one matrix approximations. Our discussion
focuses on obtaining
the first pair of components. Subsequent pairs of components
can be obtained by applying the method sequentially on the residuals
from lower-rank approximations.

The first pair of singular vectors of a data matrix
$\mathbf{X}=(x_{ij})_{m\times n}$ can be obtained by solving a least
squares problem as
\[
(\widehat{\mathbf{u}}, \widehat{\mathbf{v}}) =\mathop{\operatorname
{argmin}}_{(\mathbf{u}, \mathbf{v})}
\bigl\|\mathbf{X}-\mathbf{u}\mathbf{v}^T\bigr\|_F^2,
\]
where $\mathbf{u}$ and $\mathbf{v}$ are $m\times1$ and $n\times1$ vectors,
respectively, and $\|\cdot\|_F$ is the Frobenius norm of a
matrix. For two-way functional data, the RSVD of
\citet{huang2009analysis} defines the regularized singular
vectors as
%
%
\begin{equation}
\label{SSVDdefinition} (\widehat{\mathbf{u}}, \widehat{\mathbf
{v}})=\mathop{
\operatorname{argmin}}_{(\mathbf{u}, \mathbf{v})} \bigl\{\bigl\|\mathbf
{X}-\mathbf{u}
\mathbf{v}^T\bigr\|_F^2+\mathcal{P}_{\bolds
\lambda}(
\mathbf{u}, \mathbf{v})\bigr\},
\end{equation}
where $\mathcal{P}_{\bolds{\lambda}}(\mathbf{u}, \mathbf{v})$
is a
regularization penalty and $\bolds\lambda$ is a vector of
regularization parameters. \citet{huang2009analysis} suggested
to use the following specific form of the penalty function:
%
%
\begin{equation}
\label{eqpenalty} \mathcal{P}_{\bolds\lambda}(\mathbf{u},
\mathbf{v}) =
\lambda_\mathbf{u} \mathbf{u}^T\bolds{\Omega}_\mathbf{u}
\mathbf{u}\cdot\|\mathbf{ v}\|^2 + \lambda_\mathbf{v}
\mathbf{v}^T\bolds{\Omega}_\mathbf{v} \mathbf{v}\cdot\|
\mathbf{u}\|^2 + \lambda_\mathbf{u} \mathbf{u}^T
\bolds{\Omega}_\mathbf{u} \mathbf{u}\cdot\lambda_\mathbf{v}
\mathbf{v}^T\bolds{\Omega}_\mathbf{v} \mathbf{v},\hspace*{-25pt}
\end{equation}
where $\bolds{\Omega}_\mathbf{u}$ and $\bolds{\Omega}_\mathbf{v}$ are symmetric and
nonnegative definite penalty matrices
that apply, respectively, to the left and right singular vectors, and
$\|
\cdot\|$ is the Euclidean norm.
The usual roughness penalties used in nonparametric smoothing literature
can be adopted to define the penalty matrices
[e.g., \citet{green1994nonparametric}]. This penalty function enjoys several
desirable properties:
(i) Invariance under scale transformations
$\bu\mapsto c \bu$ and $\bv\mapsto\bv/c$ for some positive constant
$c$;
(ii) Equivariance under rescaling of $\bX$ and the fit $\bu\bv^T$;
(iii) For $\bOmega_u =0$, the penalty specializes to the one-way
penalty of \citet{silv96} for functional PCA.
See \citet{huang2009analysis} for more discussions. 

To achieve robustness, we replace the squared-error loss
in \eqref{SSVDdefinition} with a robust loss function.
Let $\rho(z)$ be a nonnegative, symmetric
function that is increasing in~$|z|$. With a slight abuse of notation,
we also use $\rho(\cdot)$ to denote the summation over elementwise
applications when the scalar function $\rho(\cdot)$ is applied to a matrix.
A~general loss function for rank-one approximation of the matrix
$\mathbf{X}$ can be written as
\[
\rho\biggl(\frac{\mathbf{X}-\mathbf{u}\mathbf{v}^T}{\sigma}
\biggr) =\sum_{i=1}^m
\sum_{j=1}^n\rho\biggl(\frac{x_{ij}-u_iv_j}{\sigma}
\biggr),
\]
where $\sigma$ is a scale parameter measuring the variability in the
approximation errors. For RobRSVD,
we define the first pair of singular vectors as
\[
(\widehat{\mathbf{u}}, \widehat{\mathbf{v}}) = \mathop{\operatorname
{argmin}}_{(\mathbf{u}, \mathbf{v})}R(
\mathbf{u}, \mathbf{v}),
\]
where
%
%
\begin{equation}
\label{robfsvddefinition} R(\mathbf{u}, \mathbf{v})=\rho\biggl
(\frac{\mathbf{X}-\mathbf{u}{\mathbf{v}}^T}{\sigma}
\biggr) +\mathcal{P}_{\bolds\lambda}(\mathbf{u}, \mathbf{v})
\end{equation}
and $\mathcal{P}_{\bolds\lambda}(\mathbf{u}, \mathbf{v})$ is the
penalty function defined in \eqref{eqpenalty}.
The determination of
the scale parameter $\sigma$ will be discussed later in
Section~\ref{secsigma}.

Our implementation uses the following
Huber's function in defining $R({\bu}, {\bv})$:
\[
\rho_\theta(x)=\cases{ %
x^2, &\quad
$\mbox{if } |x|\leq\theta,$
\vspace*{2pt}\cr
2\theta|x|-\theta^2, &\quad $ \mbox{if } |x|>\theta,$}
\]
where $\theta$ is a parameter that controls the robustness level and
a smaller value of $\theta$ usually leads to more robust estimation.
Our implementation uses $\theta=1.345$, the value commonly used in
robust regression
that produces 95\% efficiency for normal errors [\citet{huberRonchetti09}].
Our numerical studies suggested that the RobRSVD is not very sensitive
to the choice of $\theta$. Instead of the Huber function, other robust
loss functions can be used as well, for example, the $L_1$ loss which
gives similar estimates. We choose the Huber function due to its easier
implementation and faster computation.

\subsection{Iterative reweighted penalized least squares
algorithm}
\label{IRLSSec}

Although $\rho(\cdot)$ is a convex function, $R(\bu, \bv)$ is
not convex with respect to the pair $(\bu, \bv)$ and, thus,
simultaneous optimization of $R(\bu, \bv)$ over $\mathbf{u}$ and
${\mathbf{v}}$ is complicated. Note that, conditional on either $\bu$ or
$\bv$, $R(\bu, \bv)$ becomes a convex function of the other
variable. This naturally suggests an iterative reweighted (penalized)
least squares (IRLS) algorithm that alternately updates $\mathbf{u}$
and $\mathbf{v}$, assuming that the penalty parameters
$\lambda_\mathbf{u}$ and $\lambda_\mathbf{v}$ are fixed values.
This section gives the details of the algorithm,
while the choice of penalty parameters will be discussed later in
Section~\ref{GCVSec}.

For notational simplicity we assume $\sigma=1$, since otherwise
$\sigma$ can be absorbed into $\rho(\cdot)$.
Let $u_i$ denote the $i$th element in $\mathbf{u}$, and $v_j$ denote
the $j$th element of $\bv$. Let $\mathbf{x}_j$ denote the $j$th
column, and $\mathbf{ x}^{(i)}$ denote the $i$th row of $\mathbf{X}$.
Let $\operatorname{Svec}(\mathbf{X})=(x_{11}, x_{21}, \ldots, x_{m1}, x_{12}, \ldots,
x_{mn})^T$ be the column vector that is obtained by stacking the
columns of $\bX$. Furthermore, let $\psi(x)=\rho'(x)$,
$W(x)=\psi(x)/x$, and $\mathbf{W}=(w_{ij})$, where
$w_{ij}=W(x_{ij}-u_iv_j)$.

Now we consider optimization of $R(\mathbf{u}, \mathbf{v})$
over $\mathbf{v}$ given $\mathbf{u}$. Taking the derivative of
$R(\mathbf{u}, \mathbf{v})$ in \eqref{robfsvddefinition} with
respect to $v_j$, we have
%
%
\begin{equation}
\label{robfsvdderivative} \frac{\partial R}{\partial v_j} =\sum_{i=1}^m
w_{ij}(x_{ij}-u_iv_j)
(-u_i)+\frac{\partial
\mathcal{P}_\lambda(\mathbf{u}, \mathbf{v})}{\partial v_j},
\end{equation}
where
\[
\frac{\partial\mathcal{P}_\lambda(\mathbf{u}, \mathbf
{v})}{\partial
\mathbf{v}}= 2 \bigl\{\mathbf{u}^T(I+\lambda_\mathbf{u}
\Omega_\mathbf{u})\mathbf{u}(I+\lambda_\mathbf{v}
\Omega_\mathbf{v})-\mathbf{u}^T\mathbf{u}\bigr\} \mathbf{v}.
\]
The root of ${\partial R}/{\partial v_j}=0$ then gives us the optimizer
with respect to $v_j$.

Let
\[
\mathcal{Y}=\operatorname{Svec}(\mathbf{X})= \pmatrix{ \mathbf{x}_1 \vspace*{2pt}
\cr
\mathbf{x}_2 \vspace*{2pt}
\cr
\vdots\vspace*{2pt}
\cr
\mathbf{x}_n},\qquad \mathcal{U}= \pmatrix{ \mathbf{u} & 0 & \cdots& 0
\vspace*{2pt}
\cr
0 & \mathbf{u} & \cdots& 0 \vspace*{2pt}
\cr
\vdots& \vdots&
\ddots& 0 \vspace*{2pt}
\cr
0 & 0 & \cdots& \mathbf{u} },
\]
$\mathcal{W}=\diag\{\operatorname{Svec}(\mathbf{W})\}$, and
$\Omega_{\mathbf{v}|\mathbf{u}}=\mathbf{u}^T(I+\lambda_\mathbf
{u}\Omega
_{ \mathbf{u}})\mathbf{u}(I+\lambda_\mathbf{v}\Omega_\mathbf
{v})-(\mathbf
{u}^T\mathbf{u})I$.
The equations ${\partial R}/{\partial v_j}=0$
lead to
\[
\mathcal{U}^T\mathcal{W}\mathcal{U}\mathbf{v}+2
\Omega_{\mathbf
{v}|\mathbf{u}}\mathbf{v}= \mathcal{U}^T\mathcal{W}\mathcal{Y}.
\]
Solving for $\mathbf{v}$, we obtain
%
%
\begin{equation}
\label{equpdate-v} \widehat{\mathbf{v}}= \bigl(\mathcal{U}^T
\mathcal{W}\mathcal{U} +2\Omega_{\mathbf{v}|\mathbf{u}}\bigr)^{-1}
\mathcal{U}^T\mathcal{W}\mathcal{Y},
\end{equation}
which is the updating formula for $\bv$ given $\bu$.
It is easy to see that this $\widehat{\mathbf{v}}$
minimizes the following penalized weighted sum of squares:
%
%
\begin{equation}
\label{robfsvddefinition1} \widetilde{R}(\mathbf{u}, \mathbf{v}) =
(\cY-
\mathcal{U}\mathbf{v})^T\mathcal{W}(\cY-\mathcal{U}\mathbf{v}) +
\mathbf{v}^T \Omega_{\mathbf{v}|\mathbf{u}}\mathbf{v}.
\end{equation}
The equation for the fitted value of $\cY$ is
\[
\widehat{\mathcal{Y}}=\mathcal{U}\widehat{\mathbf{v}}=\mathcal
{U}\bigl(
\mathcal{U}^T\mathcal{W}\mathcal{U}+ 2\Omega_{\mathbf
{v}|\mathbf{u}}
\bigr)^{-1}\mathcal{U}^T\mathcal{W}\mathcal{Y}.
\]
Equivalently, we denote $\widehat\cY= \cH\cY$ with the hat matrix
$\mathcal{H}$ defined as
\[
\mathcal{H}=\mathcal{U}\bigl(\mathcal{U}^T\mathcal{W}\mathcal{U}+ 2
\Omega_{\mathbf{v}|\mathbf{u}}\bigr)^{-1}\mathcal{U}^T\mathcal{W}.
\]

Similarly, let
\[
\mathcal{Y}^*=\operatorname{Svec}\bigl(\mathbf{X}^T\bigr)= \pmatrix{
\mathbf{x}^{(1)} \vspace*{2pt}
\cr
\mathbf{x}^{(2)}\vspace*{2pt}
\cr
\vdots\vspace*{2pt}
\cr
\mathbf{x}^{(m)}},\qquad \mathcal{V}= \pmatrix{
\mathbf{v} & 0 & \cdots& 0 \vspace*{2pt}
\cr
0 & \mathbf{v} & \cdots& 0
\vspace*{2pt}
\cr
\vdots& \vdots& \ddots& 0 \vspace*{2pt}
\cr
0 & 0 & \cdots&
\mathbf{v} },
\]
$\mathcal{W}^*= \diag\{\operatorname{Svec}(\mathbf{W}^T)\}$,
and $\Omega_{\mathbf{u}|\mathbf{v}}=\mathbf{v}^T(I+\lambda_\mathbf
{v}\Omega_{\mathbf{v}})\mathbf{v}(I+\lambda_\mathbf{u}\Omega
_\mathbf{u})-(\mathbf
{v}^T\mathbf{v})I$.
Setting ${\partial R}/{\partial u_i}=0$, we have
\[
\mathcal{V}^T\mathcal{W}^*\mathcal{V}\mathbf{u}+2
\Omega_{\mathbf
{u}|\mathbf{v}}\mathbf{u}= \mathcal{V}^T\mathcal{W}^*
\mathcal{Y}^*.
\]
Solving for $\bu$ gives the following updating formula for $\bu$ given
$\bv$:
%
%
\begin{equation}
\label{equpdate-u} \widehat{\mathbf{u}}= \bigl(\mathcal{V}^T
\mathcal{W}^*\mathcal{V} +2\Omega_{\mathbf{u}|\mathbf{v}}\bigr)^{-1}
\mathcal{V}^T\mathcal{W}^*\mathcal{Y}^*.
\end{equation}
This $\widehat{\mathbf{u}}$ also solves a penalized weighted least
squares problem and the corresponding hat matrix is
$\mathcal{H}^*=\mathcal{V}(\mathcal{V}^T\mathcal{W}^*\mathcal{V}
+ 2\Omega_{\mathbf{u}|\mathbf{v}})^{-1}\mathcal{V}^T\mathcal{W}^*$.

The IRLS algorithm takes the results from the SVD as the initial values,
and alternately applies \eqref{equpdate-v} and \eqref{equpdate-u}
until convergence. The convergence of the algorithm is guaranteed
because each iteration step reduces the objective function, which has
a lower bound. For identifiability, at the end of each iteration step,
we normalize both $\widehat{\bu}$ and $\widehat{\bv}$ to have unit
$L_2$ norm. Upon convergence, the normalizing constant obtained in the
last iteration step will be the estimate for the corresponding singular value.

Note that the weighting matrix $\bW$ needs to be
updated at each iteration.
The matrix computation in \eqref{equpdate-v}
and \eqref{equpdate-u} can be efficiently implemented using the block
diagonal structure of the matrices.
Let $\mathbf{w}_{(j)}$ be the $j$th column of $\mathbf{W}$, and
$\mathbf{w}^{(i)}$ be
the $i$th row of $\mathbf{W}$. It can be shown that
%
%
\begin{eqnarray}
\label{SimplifiedUpdating} \mathcal{U}^T\mathcal{W}\mathcal{U} &=&
\diag\biggl\{\sum_{j}\bu^T\diag(\mathbf{w}
_j)\bu\biggr\},
\nonumber\\
\mathcal{V}^T\mathcal{W}^*\mathcal{V} &=& \diag\biggl\{\sum
_{i}\bv^T\diag\bigl(\mathbf{w}^{(i)}\bigr)\bv
\biggr\},
\nonumber
\\[-8pt]
\\[-8pt]
\nonumber
\mathcal{U}^T\mathcal{W}\mathcal{Y} & =& \diag\biggl\{\sum
_j\bu^T\diag(\mathbf{w}_j)
\mathbf{x}_j\biggr\},
\\
\mathcal{V}^T\mathcal{W}^*\mathcal{Y^*} &=& \diag\biggl\{\sum
_i\bv^T\diag\bigl(\mathbf{w}^{(i)}
\bigr)\mathbf{x}^{(i)}\biggr\}.\nonumber
\end{eqnarray}
These identities help significantly simplify the matrix computation.
Moreover, sparse matrix algorithms can be applied for efficient
computation since both
$\Omega_{\mathbf{u}|\mathbf{v}}$ and $\Omega_{\mathbf{v}|\mathbf{u}}$
are banded matrices.

%

\subsection{Penalty parameter selection} \label{GCVSec}
Following \citet{huang2009analysis}, we nest penalty parameter
selection inside the alternating algorithm that optimizes
$\bu$ for fixed $\bv$, and $\bv$ for fixed $\bu$.
Let $\widehat{\bv}^*
=(\mathcal{U}^T\mathcal{W}\mathcal{U})^{-1}\mathcal{U}^T\mathcal
{W}\mathcal{Y}$
denote the unregularized update of $\bv$, that is, the update
of $\bv$ corresponding to $\lambda_\bv=0$.
The GCV criterion for selecting $\lambda_\mathbf{v}$ conditional on
$\lambda_\mathbf{u}$ is
\[
\operatorname{GCV}(\lambda_\mathbf{v}|\lambda_\mathbf{u}) =
\frac{\|\widehat{\mathbf{v}}-\widehat{\mathbf{v}}^*\|
^2/n}{(1-\operatorname{tr}(\mathcal
{H})/n)^2}.
\]
Let $\widehat{\mathbf{u}}^*=(\mathcal{V}^T\mathcal{W}^*\mathcal
{V})^{-1}\mathcal{V}^T\mathcal{W}^*\mathcal{Y}^*$
denote the unregularized update of $\bu$.
The GCV criteria for selecting $\lambda_\mathbf{u}$ conditional on
$\lambda_\mathbf{v}$ is
\[
\operatorname{GCV}(\lambda_\mathbf{u}|\lambda_\mathbf{v})=
\frac{\|\widehat{\mathbf{u}}-\widehat{\mathbf{u}}^*\|
^2/m}{(1-\operatorname{tr}(\mathcal{H}^*)/m)^2}.
\]
These GCV formulas can be derived as a modification of
appropriately defined leave-one-row/column-out cross-validation
criteria. Details of the derivation are given in Section~1 of the online
supplemental article [\citet{zhang2013online}].
We minimize the GCV criterion to select the optimal
penalty parameters, which is done by using grid search
in our implementation. Penalty parameter selection using
the GCV formulas has much less computational complexity than
directly using cross-validation.
In our numerical experiments it usually took seconds for
one entire iteration of the algorithm including penalty parameter
selection.

%

\subsection{\texorpdfstring{Estimation of $\sigma$}{Estimation of sigma}} \label{secsigma}
We have fixed the scale parameter $\sigma$ in our development so far.
In practice, $\sigma$ can be estimated from the data using residuals
from a preliminary rank-one approximation of $\bX$. Specifically,
consider the residual matrix
$R=(r_{ij}) = \mathbf{X}-\widehat{\mathbf{u}}\widehat{\mathbf
{v}}^T$, where
$\widehat{\mathbf{u}} \widehat{\mathbf{v}}^T$ is a rank-one matrix.
The normalized Median Absolute Deviation (MAD), defined as
%
%
\begin{equation}
\label{sigmaestimate} \widehat{\sigma}=\frac{1}{0.675}\operatorname
{Med}_{ij}\bigl(|r_{ij}|,
r_{ij} \neq0\bigr),
\end{equation}
provides an estimate of $\sigma$ [\citet{MaronnaMartinYohai2006}].
In \eqref{sigmaestimate}, the $\widehat{\bu}$ and
$\widehat{\bv}$ can be obtained using the SVD or by minimizing
a robust loss function in rank-one approximation. We found that
using the SVD works very well and there is no need to resort to a
computationally more complicated robust loss function.
The RobRSVD procedure can also be applied iteratively, where
residuals from previous application are used to estimate the
scale parameter, but our experience suggests that such iteration
is usually not necessary. Hence, standard SVD is used to estimate
the scale parameter for our numerical studies.

%

\subsection{Missing values} \label{MissingSec}
In some situations, the data set may contain missing values, such as
the mortality data set analyzed in this paper or sparse functional data
as discussed in \citet{yao2005functional}. The IRLS algorithm
can still be applied with some slight modification on the updating equations
\eqref{equpdate-v} and \eqref{equpdate-u}. One approach is
to redefine $\cY$, $\cU$, $\cW$, $\cY^*$, $\cV$ and $\cW^*$
by removing the rows/columns of these matrices that contain the missing entries.
However, this approach is computationally inefficient, since
the calculation of $\cU^T\cW\cU$ and $\cV^T\cW^*\cV$ cannot
be simplified as in \eqref{SimplifiedUpdating}.

Below we develop a more efficient algorithm to deal with missing
entries. We propose to
iteratively impute the missing values and then apply the IRLS algorithm.
Each missing entry $X_{ij}$ is replaced by $\widehat{u}_i\widehat{v}_j$,
where $\widehat{u}_i$ and $\widehat{v}_j$ are obtained from
the previous iteration. The initial round of imputation can
use either the row-wise mean of the nonmissing entries in the same
row or the column-wise mean of the nonmissing entries in the same
column. Our experience suggests that both initialization methods
lead to the same results at convergence. Our proposed imputation
approach can be thought as an application of the MM
algorithm [\citet{huntlang04}], which has nice convergence properties;
see Section~2 of the online supplemental article for details [\citet
{zhang2013online}].
Similar iterative imputation approaches have been used in the literature;
see, for example, \citet{br03}, \citet{martinez09} and \citet{Lee10}.

\subsection{Function space view}
\label{secRKHS}

So far our formulation of RobRSVD is in finite dimensions, although the use
of regularization penalties implicitly assumes that there are
underlying smooth functions. We now use the
Reproducing Kernel Hilbert Space (RKHS) theory to extend our
formulation to function spaces. We refer to a standard reference
such as \citet{wahb90} for the necessary
background.

We assume $\bX= (X(y_i,z_j))_{i=1,\ldots,n; j=1,\ldots,m}$
contains the evaluations of a realization of a random field $X(y,z)$ at
$(y_i, z_j)$, where $y_i$ and $z_j$ are distinct sampling points in
the respective domains $\cY$ and $\cZ$. Seeking a rank-one
or product approximation $X(y,z) \simeq U(y) V(z)$ in function
spaces, we assume that $U(y)$ and $V(z)$ are members of
RKHSs $\cH_u$ and $\cH_v$ defined, respectively, on
the domains $\cY$ and~$\cZ$.
The RKHSs carry reproducing kernels $K_u(y_1,y_2)$ and $K_v(z_1,z_2)$,
inner products $\langle U_1, U_2 \rangle_u$ and $\langle V_1, V_2
\rangle_v$, as well as norms $\| U \|_u$ and $\| V \|_v$,
respectively.
For arbitrary $\bu=(u_1,\ldots,u_n)^T \in\bbR^n$ there is
a unique $U \in\cH_u$ interpolating $\bu$, that is, satisfying $u_i =
U(y_i)$ $(i=1,\ldots,n)$ and having minimum norm $\| U \|_u$
among all interpolants. Moreover,
this function is of the form $U(y) = \sum_{i=1,\ldots,n} c_i
K_u(y_i,y)$, and $\|U\|_u^2 = \bu^T \bOmega_u \bu$, where $\bOmega
_u =
\bK_u^{-1}$ and $\bK_u = (K_u(y_{i'},y_{i''}))_{i',i''=1,\ldots,n}$.
The same argument yields $V(z) = \sum_{j=1,\ldots,m} d_j K_v(z_j,z)$
for given $\bv\in\bbR^m$, and $\bOmega_v = \bK_v^{-1}$.
The function space version of the criterion $R(\bu,\bv)$
\eqref{robfsvddefinition} is (with some abuse of notation)
%
%
\begin{eqnarray}
\label{eqinterp-crit} R(U,V) & =& \rho\biggl(\frac{{\bX}-{\bu}{\bv
}^T}{\sigma} \biggr) +
\lambda_{u} \|U\|_u^2 \|\bv\|^2
\nonumber
\\[-8pt]
\\[-8pt]
\nonumber
&&{} + \lambda_{v} \|\bu\|^2 \|V\|_v^2
+ \lambda_{u}\|U\|_u^2 \cdot
\lambda_{v} \|V\|_v^2,
\end{eqnarray}
where $\bu= (U(y_1), \ldots, U(y_n))^T$ and $\bv= (V(z_1), \ldots,
V(z_m))^T$.

The representer theorem argument
[\citet{kimewahb71}] shows that minimization
of $R(U,V)$ in the RKHSs can be reduced to minimization of
$R(\bu,\bv)$ in the finite-dimensional space. Specifically,
if $\tilde\bu$ and $\tilde\bv$ are minimizers of $R(\bu,\bv)$, and
$\tilde{U}$ and $\tilde{V}$ are their unique interpolants
in RKHSs $\cH_u$ and $\cH_v$,
then $\tilde{U}$ and $\tilde{V}$ are the minimizers of $R(U,V)$.
This result suggests that our methodological discussions in finite-dimensional
space are without loss of generality. An important application of this result,
however, is that it allows us to extend the output vectors
$\tilde\bu$ and $\tilde\bv$ to their function space counterparts
$\tilde{U}$ and $\tilde{V}$ through the RKHS interpolation.

In the nonparametric smoothing literature, an integrated squared second
derivative penalty is commonly used. Applying this penalty to our
setting means using
$\|U\|_u^2 = \int\{U''(y)\}^2 \,dy$ and $\|V\|_v^2 = \int\{U''(z)\}^2
\,dz$
in \eqref{eqinterp-crit}. The corresponding RKHSs
$\cH_u$ and $\cH_v$ are Sobolev spaces of functions with reproducing
kernels defined in Chapter~1 of \citet{wahb90}. On the other hand,
for this special kind of penalty, we do not need the machinery
of RKHS for connecting finite-dimensional and functional spaces.
We can resort to the standard results of natural cubic
splines [see Chapter~1 of \citet{green1994nonparametric}].
There are closed-form expressions of the penalty matrices in terms
of the evaluation points and interpolation formulas available;
see Section~5 of \citet{huang2008functional}.

\section{Simulation studies} \label{SimuSec}
Three simulation studies were conducted to compare the performance of
RobRSVD against the standard SVD and the RSVD
of \citet{huang2009analysis}. The underlying true signal matrix was
generated to be either rank one, or rank one with missing values, or rank
two. A detailed analysis of the rank-one signal matrix is reported in
Section~\ref{subsecrank1}. To save space, we only summarize the
findings for the other two settings in Sections~\ref{subsecmissing}
and \ref{subsecrank2}, and present details of the studies in Section~3 of
the online supplemental document [\citet{zhang2013online}].

\subsection{Rank-one signal matrix}\label{subsecrank1}
\begin{figure}

\includegraphics{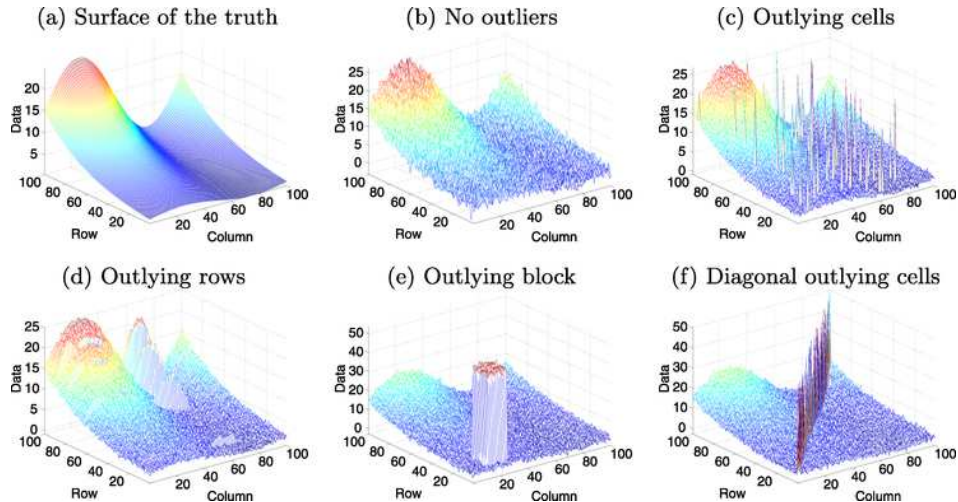}

\caption{Rank-one simulation: The surface plots. \textup{(a)} No
noise and no outliers,
\textup{(b)} no outliers with noise, \textup{(c)} random outlying
cells, \textup{(d)} outlying rows,
\textup{(e)} outlying blocks, and \textup{(f)} diagonal outlying
cells.}\label{meshplot4allsimulations}\vspace*{-3pt}
\end{figure}

We consider the following rank-one two-way functional model:
%
%
\begin{equation}
\label{eqnsimu} X(y,z)=s_0 u_0(y) v_0(z)+
\varepsilon(y,z),
\end{equation}
where $s_0=773$ is a scalar, and the two functions are $u_0(y)=(\log
10/9)10^y$ and $v_0(z)=(1+1/\pi)^{-1}\sin(2\pi z)$ with $y\in[0, 1]$,
$z\in[0, 1]$. Note that \eqref{eqnsimu} is slightly different
from\vadjust{\goodbreak}
the general two-way functional model \eqref{eqfunc-svd} in that the
two functions are now normalized: $\int_0^1u_0^2(y)\,dy=1$ and $\int
_0^1v_0^2(z)\,dz=1$, which makes it necessary to have the scalar $s_0$.
To simulate the functional data matrix, we consider 100 equal-spaced
grids in either direction. The true two-way signal surface without any
noise is plotted in panel (a) of Figure~\ref{meshplot4allsimulations}.

%
%

As a benchmark scenario, we consider the situation where the data have
no outliers. In addition, we study four different scenarios that
outliers can occur in two-way functional data: (1) random outlying
cells, (2) outlying rows, (3) outlying blocks, and (4) diagonal outlying
cells. Under each setting, the outliers are introduced as discussed
below. Besides the outliers, independent Gaussian noises $\varepsilon
(y,z)$ with mean 0 and variance $\sigma^2$ are added to the simulated
data. We consider different variances: $\sigma^2=0.2, 0.5, 0.8, 1$. For
each simulation setting, 100 simulation replications are performed. The
surface plot of one random replication (with $\sigma^2=1$) is plotted
in Figure~\ref{meshplot4allsimulations} for each of the four outlying
scenarios, respectively.

We now describe how the outliers are introduced for each simulation
setting. Let $\mathbf{X}_0=s_0\mathbf{u}_0\mathbf{v}_0^T$ denote the
signal matrix (i.e., without any noise), where $\mathbf{u}_0$ (or
$\mathbf{v}_0$) denotes the vector that contains the observed values of
the function $u_0(y)$ [or $v_0(z)$] at the 100 equally-spaced grid
points within $[0,1]$: 
\begin{longlist}
\item[1.]\textit{Outlying cells}: Under this setting, we randomly select 100
cells in the data and replace their entries with outlying values. In
particular, the values in the selected cells are randomly simulated
from the uniform distribution with support $[C_1, 2C_1]$ with $C_1=\max
(\bX_0)$. 
%
\item[2.]\textit{Outlying rows}: We randomly select five rows, and replace
them by five new rows defined below.\vadjust{\goodbreak} For each of the five randomly
selected rows, we obtain the outlying curve by multiplying the
corresponding $s_0u_0(y)$ with a different function ${v}_1(z)=C(1+\sin
(4\pi z))$ with $C$ being the normalizing constant. Note that the curve
shapes of the outlying rows are different from the shape of the other rows.
\item[3.]\textit{Outlying block}: We randomly select a continuous square
block of cells at a randomly selected location, with the block size
fixed as $10\times10$. Within the block, we shift the cells upward by
adding a random amount, which is uniformly distributed on $[2C_1,
3C_1]$. 
%
\item[4.]\textit{Diagonal outliers}: 
We replace the diagonal entries of the matrix with values uniformly
distributed between $[C_1, 2C_1]$. This setting mimics the cohort
effects observed in the Spanish mortality data (Section~\ref{spaindataanalysis}).
\end{longlist}

%
%
\begin{figure}

\includegraphics{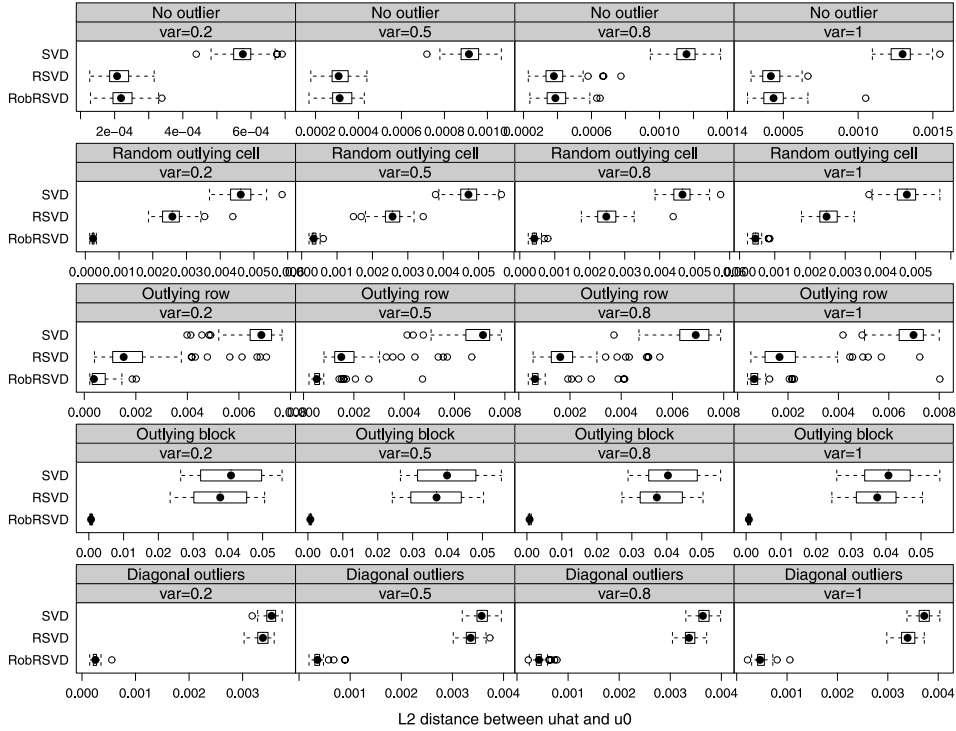}

\caption{Rank-one simulation:
Boxplots of the $L_2$ distance between $\widehat{\mathbf{u}}_0$ and
$\mathbf{u}_0$.} \label{l2dist4uplot}
\end{figure}

%
%
\begin{figure}

\includegraphics{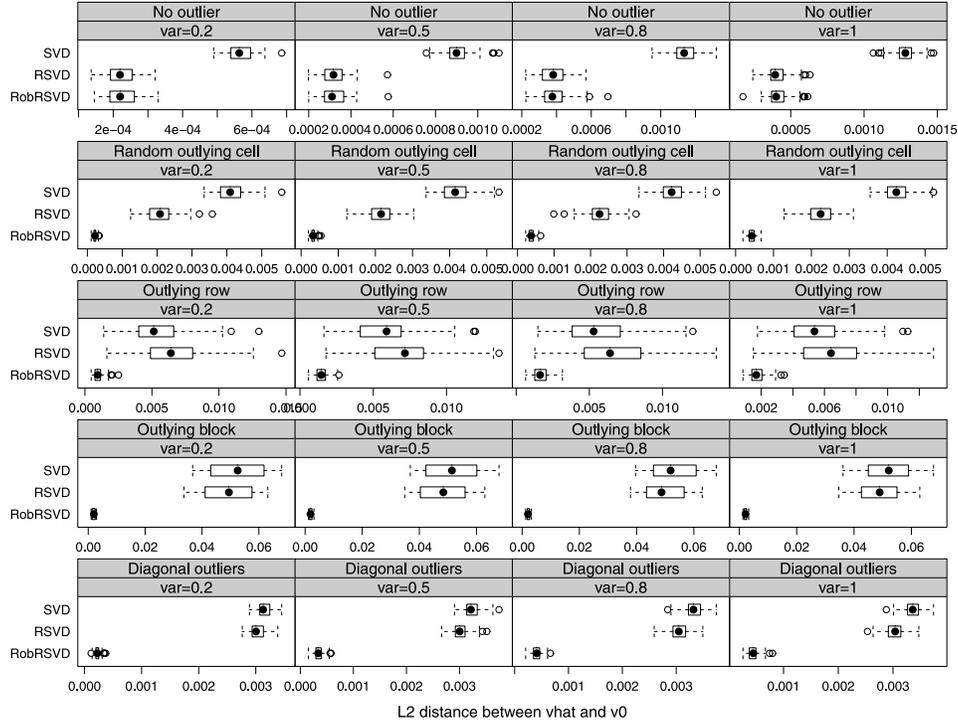}

\caption{Rank-one simulation: Boxplots of the $L_2$ distance
between $\widehat{\mathbf{v}}_0$ and $\mathbf{v}_0$.} \label{l2dist4vplot}
\end{figure}

The three methods, SVD, RSVD and RobRSVD, were applied to the 100
simulated data sets under each setting, and the best rank-one
approximations were obtained to get the estimates for $\mathbf{u}_0$
and $\mathbf{v}_0$. The penalty parameters of the RSVD and RobRSVD were
selected using the GCV method.

To compare various methods, we calculated the $L_2$ distance between
the estimates and the truth for each simulated
data. Figures~\ref{l2dist4uplot} and \ref{l2dist4vplot} present the
boxplots of the 100 distances for the three methods for $\mathbf{u}_0$
and $\mathbf{v}_0$, respectively, for each of the four noise levels
and each of the outlier scenarios.

In summary, both figures clearly show that:
\begin{longlist}[1.]
\item[1.] For the benchmark no-outlier cases, RobRSVD and RSVD perform
comparably, and both are better than SVD due to smoothing
regularization; this suggests that our RobRSVD does not lose much when
the data contain no outliers.
\item[2.] For all the outlying settings, RobRSVD improves significantly
over RSVD and SVD, which supports the robustness of RobRSVD against
various kinds of outliers in two-way functional data. RobRSVD has the
smallest median $L_2$ distance and variability across all the settings
and the different noise levels.
\end{longlist}
%

%
%
\begin{figure}

\includegraphics{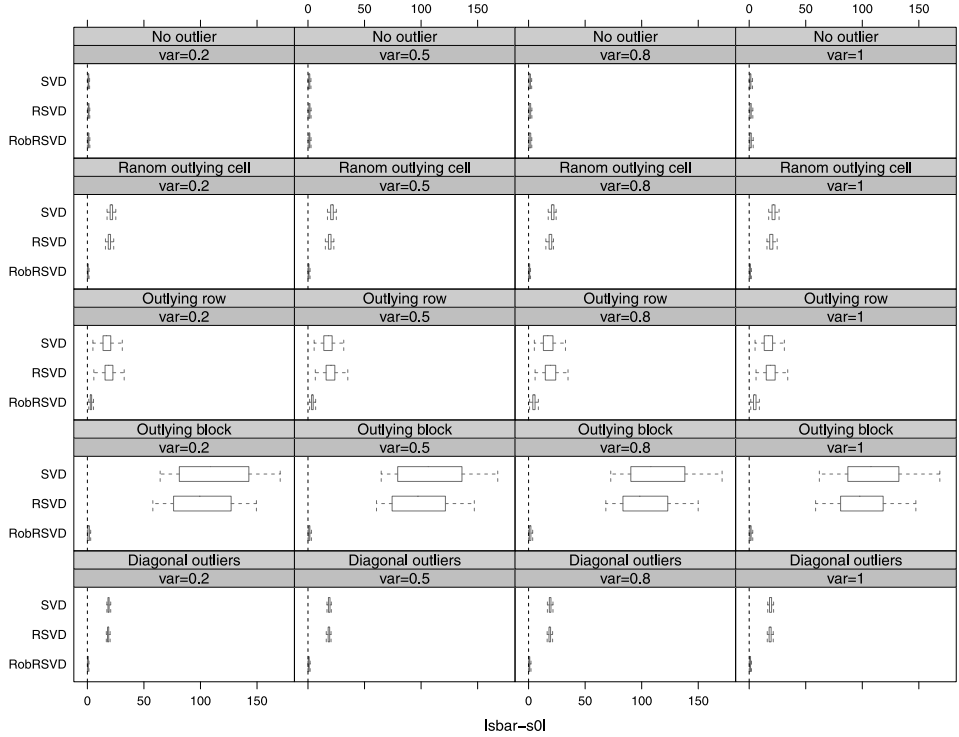}

\caption{Rank-one simulation: Boxplots of $|\widehat{s}_0 -
s_0|$.} \label{dist4splot}
\end{figure}

We also calculated the estimated singular values $\widehat{s}_0$ and
compared them with the true singular value $s_0=773$. For each method
and each noise level, Figure~\ref{dist4splot} presents the boxplot of
the 100 absolute differences between $\widehat{s}_0$ and $s_0$. The
comparison shows that RobRSVD performs similarly with SVD and RSVD for
cases with no outliers, while much better when there are outliers.

%
%
\begin{figure}

\includegraphics{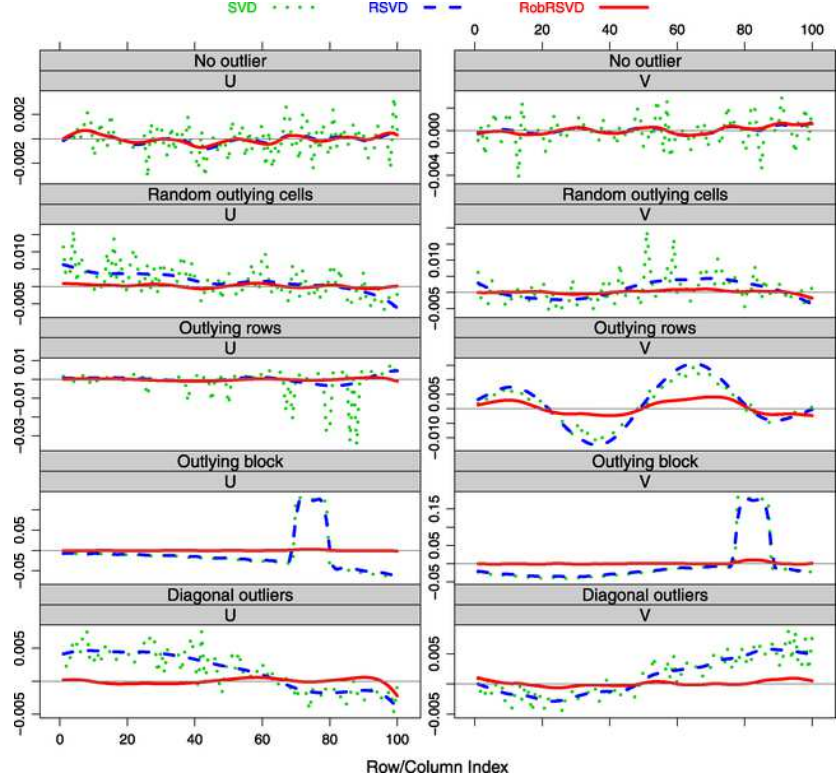}

\caption{Rank-one simulation: Comparison of individual
estimates obtained
from the data plotted in Figure \protect\ref{meshplot4allsimulations}.
The differences between the estimates and
the truth are plotted. The RobRSVD method shows the most robust
performance.} \label{UVDiffplot4AllCases}
\end{figure}

To get some ideas about individual estimation performance, Figure~\ref{UVDiffplot4AllCases} compares the estimates obtained from the
particular data sets shown in 
Figure~\ref{meshplot4allsimulations}, by plotting the differences
between the estimated curves and the true curve [either $u_0(\cdot)$ or
$v_0(\cdot)$]. As one can see, the RobRSVD method is again the clear
winner. We also observe that the smoothing step in RSVD can mitigate
the outlying effects to some extent in certain cases, but still cannot
fully remove those effects. The additional incorporation of robust loss
function in RobRSVD further improves the robustness of RSVD.

\subsection{Rank-one signal matrix with missing values}\label{subsecmissing}
Our motivating Spanish mortality data contain both outliers and
missing values, which motivates us to investigate the performance of
RobRSVD when there are missing values. For each simulated data set
considered in Section~\ref{subsecrank1}, we randomly selected and
deleted 100 cells from it to form a new data set with missing values.
We used the imputation method described in Section~\ref{MissingSec}
to estimate $\mathbf{u}$ and $\mathbf{v}$ for SVD, RSVD and RobRSVD.

The simulation results are reported in the online
supplement. The comparison presented in
Figures~1 and 2 there clearly shows that the RobRSVD remains to
be the winner across all the settings considered.

\subsection{Rank-two signal matrix}\label{subsecrank2}
We also studied the situation where the true signal matrix is rank
two, using a setting similar to what has been studied by
\citet{huang2009analysis}.
Similar to Section~\ref{subsecrank1}, we considered five simulation
scenarios: no outliers, outlying cells, outlying rows, outlying block,
and diagonal outliers. Detailed descriptions can be found in the online
supplement, with comparative results presented in Figures~3--5 there.

We used two measures to gauge\vspace*{1pt} the performance of estimating the rank-2
signal matrix. The first measure is $\|\widehat{\mathbf{X}}_0-\mathbf
{X}_0\|_F$,
the Frobenius norm of the difference between the estimated best
rank-two matrix $\widehat{\mathbf{X}}_0$ and the true signal matrix
${\mathbf{X}}_0$. The second measure the largest \textit{principal
angle} [\citet{golub1996matrix}] between the true subspace and the
subspace spanned by the corresponding singular vector estimates.
Specifically, let $\mathbf{U} =\operatorname{span}(U_1^*, U_2^*)$
denote the linear subspace spanned by $U_1^*(y)$ and $U_2^*(y)$
evaluated at the grid points and $\widehat{\mathbf{U}}$ be the
corresponding estimate of this subspace. The principal
angle between $\mathbf{U}$ and $\widehat{\mathbf{U}}$ can be computed
as $\operatorname{cos}^{-1}(\rho) \times180/\pi$, where $\rho$ is the minimum
eigenvalue of the matrix $Q^T_{\widehat{\mathbf{U}}} Q_\mathbf{U}$
where $Q_{\widehat{\mathbf{U}}}$ and $Q_\mathbf{U}$ are orthogonal basis
matrices obtained by the QR decomposition of the matrices
${\widehat{\mathbf{U}}}$ and $\mathbf{U}$, respectively.
RobRSVD performed the best in all cases with outliers
under both distance measures, while RSVD and RobRSVD
usually performed similarly and were better than SVD
in cases without outliers.

\section{The Spanish mortality data}\label{spaindataanalysis}
In this section we analyze the Spanish mortality data using various
methods to illustrate the benefits of our proposed RobRSVD method. The
Spanish mortality data are available in the Human Mortality Database
[\citet{HMD}]. This mortality data set was collected such that each row
represents a year between 1908 and 2007, each column represents an age
group from 0 to 110, and each cell records the mortality rate for a
particular age group during that year. The data are naturally two-way
functional, since each column vector is a time series of mortality rate
of a given age group, and each row vector is a mortality curve of
different age groups at a specific year.

\citet{zhang2007singular} developed several visualization tools for
exploring two-way functional data, which were used to analyze a subset
of the Spanish mortality data. As a result, they identified a couple of
interesting outlying time periods (i.e., rows in the data matrix):
\begin{itemize}
\item the 1918 Spanish flu pandemic, and
\item the 1936--1939 Spanish Civil War,
\end{itemize}
both of which experienced the death of millions of Spanish people (in
an unusual
age distribution). In both cases, the mortality rate increased well
above what the normal yearly trend would have predicted, and the
authors noted that the outlying years affected the estimation of the
first few leading SVD components, which is consistent with our findings
reported below.

One can view the mortality rate data as some normal mortality trend, a
function of age group and year, contaminated with additive noises,
including measurement errors and potential outliers. Hence, a good
estimation method should be able to recover the underlying normal
mortality varying pattern across age and year, with minimal effects of
the noises including the outliers.

Before the formal analysis, we make two comments regarding the data.
Following \citet{zhang2007singular}, the data were first transformed
through $\log_2(X+1/2)$ where $X$ denotes the original mortality rate.
There are missing values for the elder people in the data, and we
employ the procedure discussed in Section~\ref{MissingSec} to
automatically accommodate the missing values.


%
%
\begin{figure}

\includegraphics{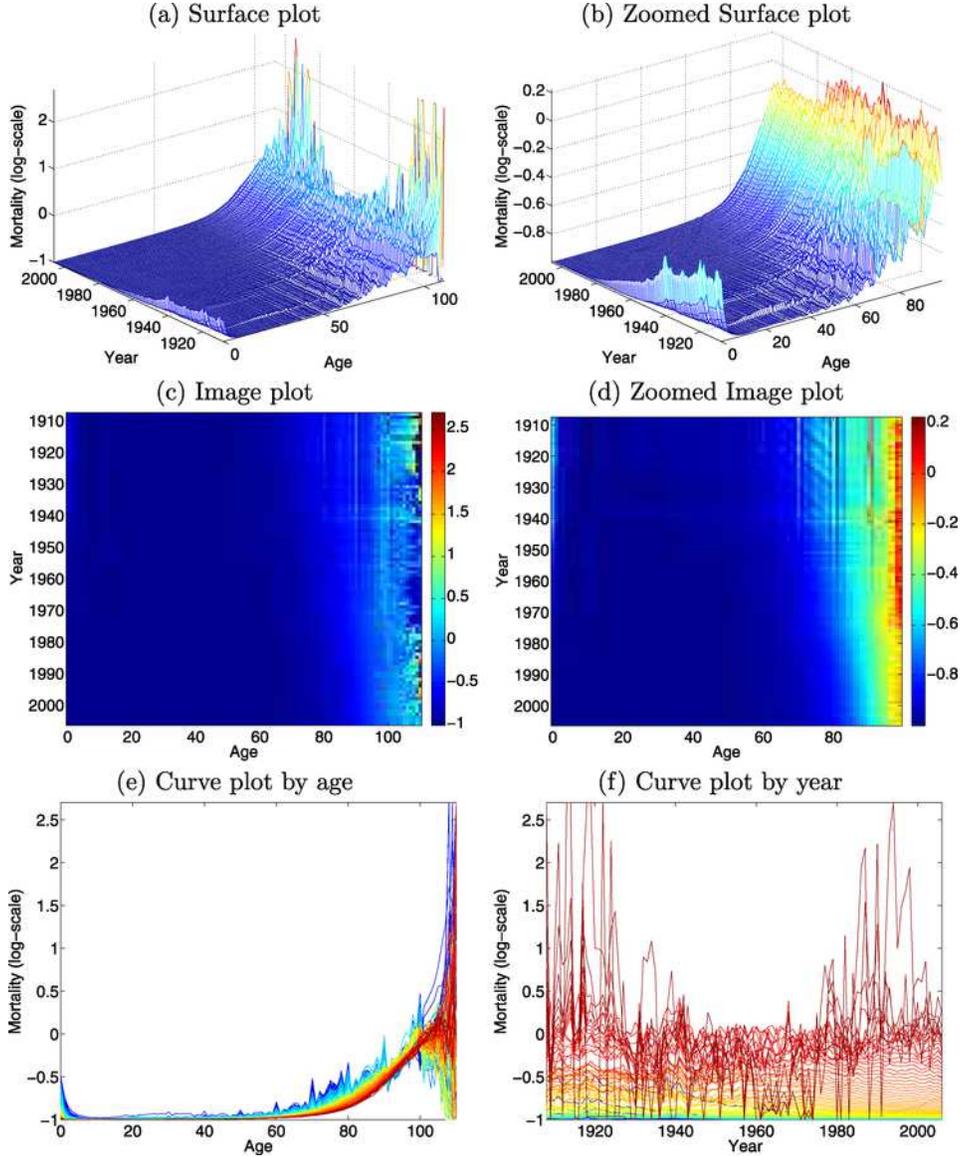}

\caption{Various visualizations of the mortality data (log-scale):
\textup{(a)} the mesh surface plot, \textup{(b)} the zoomed surface
plot (up to age 100),
\textup{(c)} the image plot, \textup{(d)} the zoomed image plot (up
to age 100),
\textup{(e)} the curve plot versus age, \textup{(f)} the curve plot
versus year.}\label
{esmortoriginalplot}
\end{figure}

Figure~\ref{esmortoriginalplot} provides several functional views of
the log-transformed data. Several interesting observations can be made
from the plots. The mesh surface plot in panel~(a) highlights the high
mortality rates among the seniors that are older than~100. 
To better depict the mortality trend among people less than 100 years
old, the zoomed surface plot in panel (b) shows the mortality rate
pattern up to age 100: for a given year, the mortality rate generally
decreases from infants to teenagers and adults younger than 60, and
begins to increase when the age is over 60, which is the standard
mortality pattern across age; for a given age group, the mortality rate
decreases across the years, which reflects the improvement of life
quality and health care; in addition, the decrease-across-year among
younger people is more significant than for elder people. For the
(zoomed) image plots on panels (c)--(d), we observe the cohort effects
discussed by \citet{zhang2007singular} showing up as the diagonal strips
and, more importantly, the two outlying time periods appearing as
horizontal strips: the 1918 flu pandemic affects all age groups, while
the 1936--1939 civil war affects only those older than 20. The curve
plots in panel (e) show the mortality rate as a function of age where
each curve corresponds to a particular year, and in panel (f) show the
mortality rate as a function of year where each curve is for a
particular age. 

To better understand the dominating modes of variation within the data,
we use SVD, RSVD and RobRSVD to find (smooth) low-rank approximations
for the data and compare their results.
Let $s_i$ be the $i$th singular value for the standard SVD.
The ratio of $s_i^2$ over the Fronbenius norm of the data matrix represents
the percentage of energy explained by the $i$th component.
The percentage can be plotted in a scree plot as a useful visual
aid for deciding the number of significant components.
For the mortality data, the scree plot based on the SVD shows a clear
knee at rank two, with the first two standard SVD components explaining
93.3\% and 5.0\% of the total energy, respectively, while the third
component accounts for less than 1.0\% of the total energy. Thus, we only
look at the first two dominating pairs of functional components
when we compare different methods.

\begin{figure}

\includegraphics{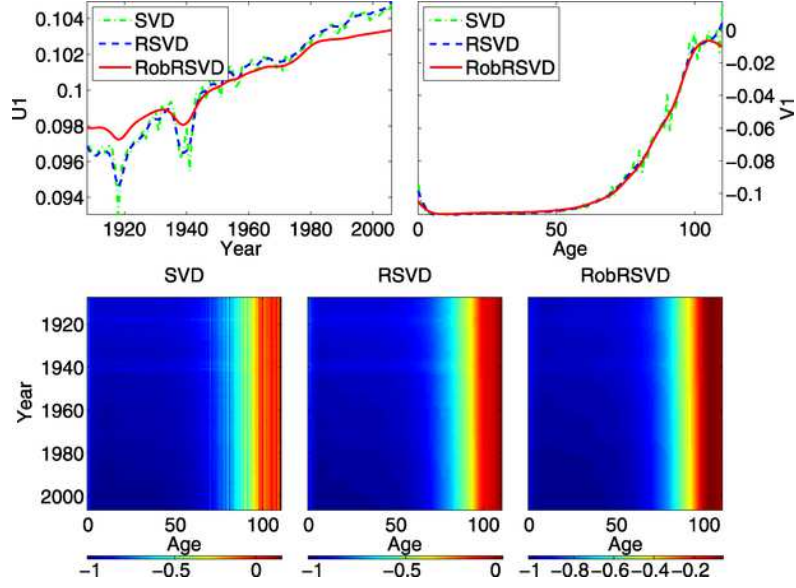}

\caption{Comparison of the first pairs of (regularized)
singular vectors.
The left (regularized) singular vectors ($\mathbf{u}_1$) from SVD and
RSVD are obviously affected
by the two outlying time periods (1918, 1936--1939).}\label{1stcomp} 
\end{figure}

Figure~\ref{1stcomp} compares the first left (regularized) singular
vectors (RSVs) ($\mathbf{u}_1$) and the first right RSVs ($\mathbf
{v}_1$), as well as the best rank-one two-way approximation from the
three methods. Note that the first pair of RSVs explains the major mode
of variation in the data. The green dotted-dash curves show the results
of the regular SVD method, the blue dash ones correspond to the RSVD
method, and the red solid curves are for our RobRSVD method. The
RobRSVD left component shows a general smooth increasing trend from
1908 to 2007, while the corresponding right component resembles the
standard smooth age-mortality curve. On the other hand, the left
functional components from SVD and RSVD are rather wiggly and seriously
affected by the two outlying time periods in 1918 and 1936--1939. 
The robustness of RobRSVD can also be seen from the image plots of the
best rank-one approximation, the bottom row of Figure~\ref{1stcomp}.
For both SVD and RSVD approximations, the outlying years show up as
horizontal strips to reflect the increased mortality rates across a
wide range of age groups. Furthermore, the RobRSVD image plot shows a
much smoother trend across age.

%
%

%
%
\begin{figure}

\includegraphics{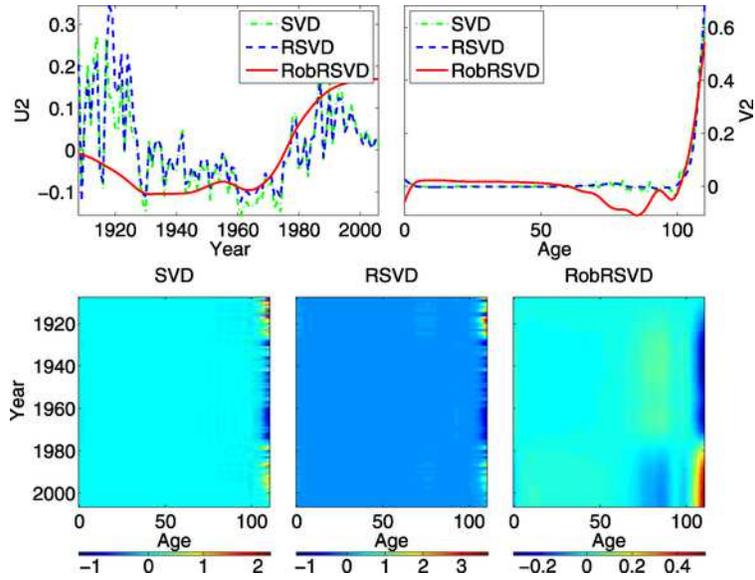}

\caption{Comparison of the second pairs of (regularized)
singular vectors.
Note the contrast in the RobRSVD image plot around 1970.} \label{2ndcomp}
\end{figure}


The second pair of (regularized) singular vectors is compared in
Figure~\ref{2ndcomp}. In general, we observe that the RobRSVD component
is smoother and more interpretable than the SVD and RSVD components,
which tend to be wiggly and show effects from the outlying years. Note
that the numerical scales of the colorbars for SVD/RSVD are much larger
than those of RobRSVD, which are caused by the outliers appearing in
the SVD/RSVD components. The second pair of the RobRSVD component
highlights the contrast between people of age 50--100 and people older
than 100 during two different time periods: before 1970, the older
group has a lower mortality rate than the younger group, while after
1970, the comparison is reversed. This contrast can be clearly seen in
the bottom right panel.


Figure~6 of the online supplement shows
the 3-dimensional surface plots of the best rank-two approximations
by the three methods, also indicating that the RobRSVD is least
influenced by outlying observations.

\section*{Acknowledgments}
We thank the Editor, the Associate Editor and the referees for
invaluable comments and suggestions, which greatly improved the quality
of this paper.

\begin{supplement}[id=suppA]
\stitle{Supplemental notes for ``Robust regularized singular value
decomposition with application to mortality data''}
\slink[doi]{10.1214/13-AOAS649SUPP} 
\sdatatype{.pdf}
\sfilename{aoas649\_supp.pdf}
\sdescription{The supplemental notes include deviation of the GCV
formula in this paper, an MM algorithm to handle missing value, two
additional simulation examples in details, and one additional plot for
the analysis of the mortality data.}
\end{supplement}

%


\printaddresses

\end{document}